\documentclass[debug]{rmaa}

\usepackage{paralist}
\usepackage{psfrag,color}
\usepackage[latin1]{inputenc}
\usepackage[dvipsnames]{xcolor}
\usepackage{ulem}

\newcommand\TA{\tablenotemark{a}}

\newcommand\zo{$O$}
\newcommand\zc{$C$}
\newcommand\msun{M$_\odot$}

\newcommand{\tc}{$t^2$}
\newcommand{\Hii}{\ion{H}{2}\null}
\newcommand{\Hiir}{\ion{H}{2} region\null}
\newcommand{\Hiirs}{\ion{H}{2} regions\null}

\title{The ADF and the \tc \ formalism in \Hii\ regions based on the upper mass limit of the IMF for the MW} 

\author{
  L. Carigi\altaffilmark{1}, 
  A. Peimbert\altaffilmark{1},
  M. Peimbert\altaffilmark{1},
  and G. Delgado-Inglada\altaffilmark{1}}

\altaffiltext{1}{Instituto de Astronom\'\i a, UNAM, Mexico.}

\shortauthor{Carigi, Peimbert, Peimbert \& Delgado-Inglada}
\shorttitle{The ADF and the \tc \ formalism in \Hii\ regions}

\fulladdresses{
\item Carigi, L., Peimbert, A., Peimbert, M. \& Delgado-Inglada, G.: Instituto de Astronom\'\i a, Universidad Nacional Aut\'onoma de M\'exico, Apdo. Postal 70-264, M\'exico, C.P. 04510, CdMx, Mexico. (carigi, antonio, peimbert, gdelgado@astro.unam.mx).}

\listofauthors{Carigi, L., Peimbert, A., Peimbert, M. \& Delgado-Inglada, G.}
\indexauthor{Carigi, L.}
\indexauthor{Peimbert, A.}
\indexauthor{Peimbert, M.}
\indexauthor{Delgado-Inglada, G.}

\abstract{
We study in depth the abundance discrepancy problem in \Hiirs, this time from a different perspective than the usual one: by studying the effect of the upper mass limit ($M_{up}$) of the initial mass function (IMF) on the O, C, and He predicted by chemical evolution models for the Milky Way. 
We use abundances determined with the direct method (DM) and with the temperature independent method (TIM). 
We compare the predicted abundances at the present time with observations of Orion, M17, and M8 to determine the $M_{up}$ value of the galactic IMF. 
From the DM abundances, the models predict an $M_{up} = 25 - 45$ \msun, 
while from the TIM, CEMs derive an $M_{up} = 70 - 110$ \msun.
Spiral galaxies with the stellar mass and star formation rate of the MW are predicted to have an $M_{up} \approx 100$ \msun. 
These results support that abundances derived from the TIM are better than those derived from the DM.
}

\resumen{
Estudiamos el problema de la discrepancia de abundancias en regiones \Hii  \ a partir del efecto
que tiene la masa superior  ($M_{up}$) de la funci\'on inicial de masa (IMF) en la producci\'on de O, C
y He obtenida por modelos de evoluci\'on qu\'imica de nuestra galaxia. 
Comparamos las abundancias determinadas por el m\'etodo directo (DM) y por el m\'etodo independiente
de la temperatura (TIM) en Ori\'on, M17 y  M8  con las abundancias modeladas actuales
para determinar el valor de la $M_{up}$ de la IMF Gal\'actica.
Utilizando las abundancias a partir del DM se requiere que $25<M_{up}<40$ \msun, 
mientras que usando las abundancias obtenidas a partir del TIM se requiere que  $70<M_{up}<110$ \msun. 
Para galaxias espirales con masa estelar y tasa de formacion estelar similar a la MW se ha predicho una $M_{up} \approx 100$ \msun.
Este resultado implica que las abundancias
obtenidas por el TIM son mas adecuadas que las obtenidas por el DM.
}

\addkeyword{H~II regions}
\addkeyword{ISM: abundances}
\addkeyword{stars: mass function}
\addkeyword{Galaxy: evolution}

\begin{document}
\maketitle

\vspace{1cm}

\section{Introduction}\label{s-intro}

The study of the chemical composition of \Hiirs\ is crucial for the understanding of the chemical composition of the universe; they can provide observational constraints required by models of galactic chemical evolution. The proper determination of the oxygen abundance in \Hiirs\ is critical to  be able to compare our determinations with other branches of astrophysics, as well as to be able to present a coherent model of the evolution of the universe.

Although a comprehensive chemical composition is desired, frequently studies are only able to determine a few chemical elements; in fact many works focus only on the oxygen abundance. A single well determined element can be quite useful, since most elements are expected to behave in an orchestrated fashion, and oxygen is the ideal candidate, since it is expected to comprise about half of the heavy element abundance, not only that, but is the easiest element to study since it is the only element that produces bright optical lines for each of its main ionic species.

There are three competing methods to derive ionic O$^+$ and O$^{++}$ abundances, relative to the H$^+$ abundance, in \Hiirs :  
i) The  use  of  [\ion{O}{2}] and [\ion{O}{3}]  collisionally excited  lines  (CELs) together with the H Balmer lines;  these lines  are relatively easy to observe, unfortunately their emmisivities depend strongly on the local temperature (in fact, this specific characteristic of CELs is used to determine the characteristic temperature of photoionized regions); this is known as the direct method (DM).
ii) The use of \ion{O}{1} and \ion{O}{2} recombination lines (RLs) together with the H Balmer lines (which are also RLs); this strategy has the advantage that the ratios of any pair of RLs are almost independent of the temperature structure, however RLs of heavy elements are quite faint and harder to work with.
iii) The use of CELs with the $t^2$ formalism introduced by \citet{peimbert67, peimbert69} in which the effect of the temperature structure on the emission lines is taken into account. Recent reviews of these methods have been presented by \citet{perez17, peimbert17, peimbert19, garcia20}.

Abundances derived from either RLs or the $t^2$  formalism usually agree with each other and are usually 0.2 to 0.3 dex higher than those determined using the DM \citep[e.g.][]{esteban09, peimbert17,peimbert19,carigi19}. We have defined the temperature independent method (TIM) as abundances derived from either RLs, the $t^2$ formalism, or their average \citep{carigi19}. On the other hand, the ratio between RL abundances to DM abundances (TIM abundances to DM abundances) is called the abundance discrepancy factor (ADF): $ADF(X^{+i}) = X^{+i}_{\rm RLs}/X^{+i}_{\rm CELs}$ \citep{tsamis03}.

Whether the origin of the abundance discrepancy is (only) due to temperature fluctuations \citep[][]{peimbert67} or chemically inhomegeneities \citep[first proposed by][]{torres90} is still under discussion \citep[see e.g.,][]{esteban18, garcia19}. Other hypothesis have not been very successful and have been proposed over the years such as the $\kappa$ distribution of electrons \citep{nicholls12} or uncertainties in the atomic data \citep[e.g.,][]{rodriguez10}, but have been discarded \citep{garcia19}. Anyway it is beyond the scope of this paper to discuss the origin of this long-standing problem.

Oxygen observations of \Hiirs\ are limited to the O$^+$ and O$^{++}$ gaseous components. In general, it is not necessary to correct for other ionization stages where  O$^{+3}$ is limited to $\sim 1$-$2\%$ and O$^0$ is considered to be outside the relevant volume of the \Hiir; and, when comparing abundances between several \Hiirs, this is enough. However, when comparing with other types of objects, or when trying to model the evolution of a galaxy, it is of critical importance to correct for oxygen atoms trapped in dust grains which are estimated to be $\sim 25\%$ of the total oxygen in \Hiirs\ \citep[i.e. an additional $\sim 35\%$, when compared to the gaseous component;][]{mesa09,peimbert10,espiritu17}. 

Here we study the discrepancy between TIM and DM abundances from a different perspective. We will compute chemical evolution models adopting the initial mass function (IMF) by \citet{kroupa02} with different $M_{up}$ values, where $M_{up}$ is the upper mass limit of the IMF. The $M_{up}$ value is not the maximum stellar mass present in a given \Hiir, but the maximum mass of the IMF averaged over the age of the Galaxy. 

In order to be able to study in depth as many details as possible, in this paper we will do a deeper study for a few of \Hiirs\ only. We have selected the Orion Nebula because it is, by far, the most studied Galactic \Hiir. For our second object we selected M17 because it is the second most studied Galactic \Hiir, it is relatively close by, yet it has an appreciably different galactocentric distance; also it is a high-ionization \Hiir\ and thus we need not worry about the possible presence of neutral helium.   Our last object is M8; it is one of the most studied Galactic \Hiirs , it has approximately the same galactocentric radius as M17. 

Chemical evolution models for the MW have been built to reproduce robust observational constraints of the Galaxy. Some authors  \citep[e.g.,][]{prantzos2018, romano2010} constrain their CEMs by trying to fit the solar abundances at the Sun's age, other authors \citep[e.g.,][]{spitoni2019, molla2015} constrain their models by trying to reproduce the [$\alpha$/Fe] - [Fe/H] trend shown by stars at the solar vicinity. Most of these models reproduce the current slope of the $\alpha$/H gradients, but the predicted absolute values are different and consequently the predicted $\alpha$ enrichment efficiencies are different during the last few Gyrs of the evolution.
 
Unfortunately, the chemical gradients of \ion{H}{2} regions can not be used as solid constraints because, while the slope of the chemical gradient is widely accepted, the absolute values, of \ion{H}{2} region gradients found in the literature, present a large dispersion; this becomes more pronounced when combined with gradients derived from other young objects.

To improve the quality of the models, it is important to fit the absolute value of the element abundances at the present time, not only the slope. These absolute values become critical to put restrictions on the CEMs. For example: since O, and other $\alpha$ elements, are mainly produced by massive stars, the absolute values of the gradient are critical to determine the $M_{up}$ value of the IMF.

For simplicity we will use H, He, C, and O to represent the abundances by number, and $X$, $Y$, $C$, and $O$ to represent the abundances by mass of these elements; $Z$ represents the total heavy element abundance by mass.

\section{H II regions Abundances and Chemical Evolution Models}\label{s-CEMs}

\subsection{O/H vs. distance to the Galactic center}

\begin{figure}
\begin{center}
\includegraphics[angle=0,scale=0.58, trim=15 25 15 10, clip]{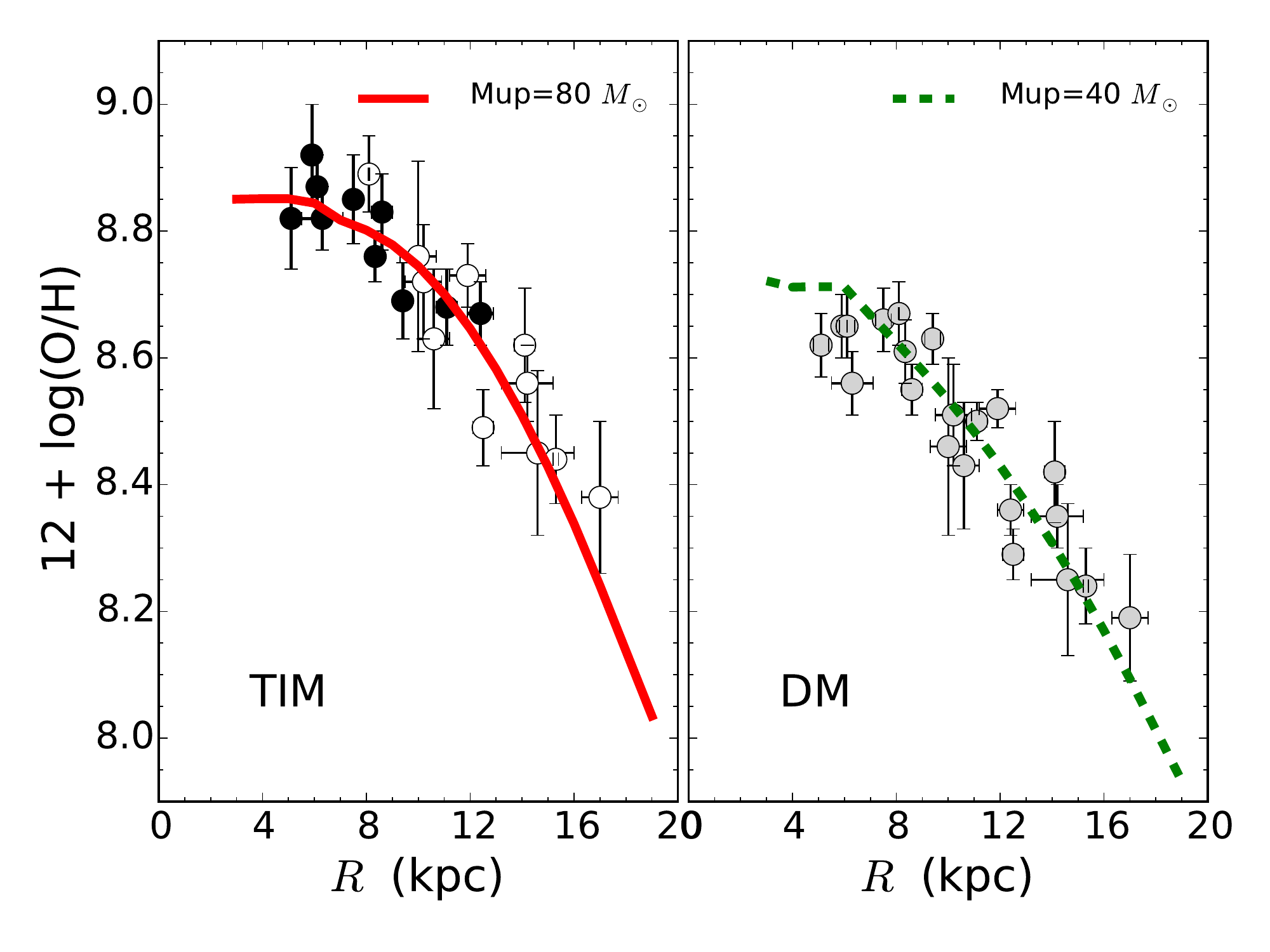}
\end{center}
\caption[f1.eps]{Values of O/H as a function of the distance to the galactic center. Models for $t^2 = 0.00$ and for observed $t^2$ values versus two chemical evolution models with different $M_{up}$ values. The filled and dashed lines represent the radial distribution obtained with the TIM ($M_{up}$ = 80 ${\rm M}_{\odot}$) and DM ($M_{up}$ = 40 ${\rm M}_{\odot}$) models, respectively. The circles represent the values derived from observations: TIM values based on recombination lines (black), TIM values based on the calibration by \citealt{pena12} (empty), and DM values (grey). For further discussion, see \citet{carigi19}. 
\label{f-O/HvsR}}
\end{figure}

In Figure~\ref{f-O/HvsR} we present two sets of {\Hiir} data for O/H, one based on the TIM and the other based on the DM, as well as the best fit models to each data set. The observational data was compiled from \citet{esteban04, esteban13, esteban16a, esteban17, fernandez17, garcia04, garcia05, garcia06, garcia07, garcia14}. The models, built to reproduce the O/H gradient, come from \citet{carigi19}. For a more detailed description of the data selection, the model, and the fit see Carigi et al.. All abundances have been corrected by the fraction of O trapped in dust grains \citep{peimbert10,pena12,espiritu17}. The figure illustrates very well that the curves required to fit the TIM and the DM data are quite different; it is thus no surprise that the $M_{up}$ required by the models to fit the TIM data and the DM data are very different: while the $M_{up}$ used to fit the TIM data amounts to 80 M$_{\odot}$, the value used to fit the DM data amounts to 40 M$_{\odot}$. 

When comparing these values with the ones observed from young objects (B-stars, Cepheids), we find that the model based on the TIM values produces an excellent fit between 5 and 17 kpc, while the model based on the DM values fails to reproduce the observations \citep{carigi19b}.

\subsection{Representation of the chemical evolution models}

We compute a set of nine chemical evolution models (CEMs) for MW like galaxies based on the work by \citet{carigi19}; these models differ only in the adopted $M_{up}$ value. We present the output of these  models for 6.2 kpc, corresponding to the average galactocentric distance of M17 (6.1 kpc) and M8 (6.3 kpc), and for 8.34 kpc, corresponding to the galactocentric distance of the Orion Nebula.

The initial abundances of our models are: $X(0) = 0.7549$, $Y(0) = 0.2451$, $C(0) = 0$, and $O(0) = 0$, where $Y(0) = 0.2451 \pm 0.0026$ is the primordial helium abundance derived by \citet{valerdi19}, their $Y(0)$ result is in good agreement with the value derived by \citet{planck18}
that amounts to $Y(0) = 0.24687 \pm 0.00076$.

We explore the $M_{up}$ effects on the predicted value of $Y$, $C$, and $O$ during the whole evolution. The galaxies are formed in an inside-out scenario of primordial infall, with the halo component formed from 0 to 1 Gyr and the disk component formed from 1 to 13 Gyr. 

\begin{figure}
\begin{center}
\includegraphics[angle=0,scale=0.55, trim= 5 25 5 10, clip]{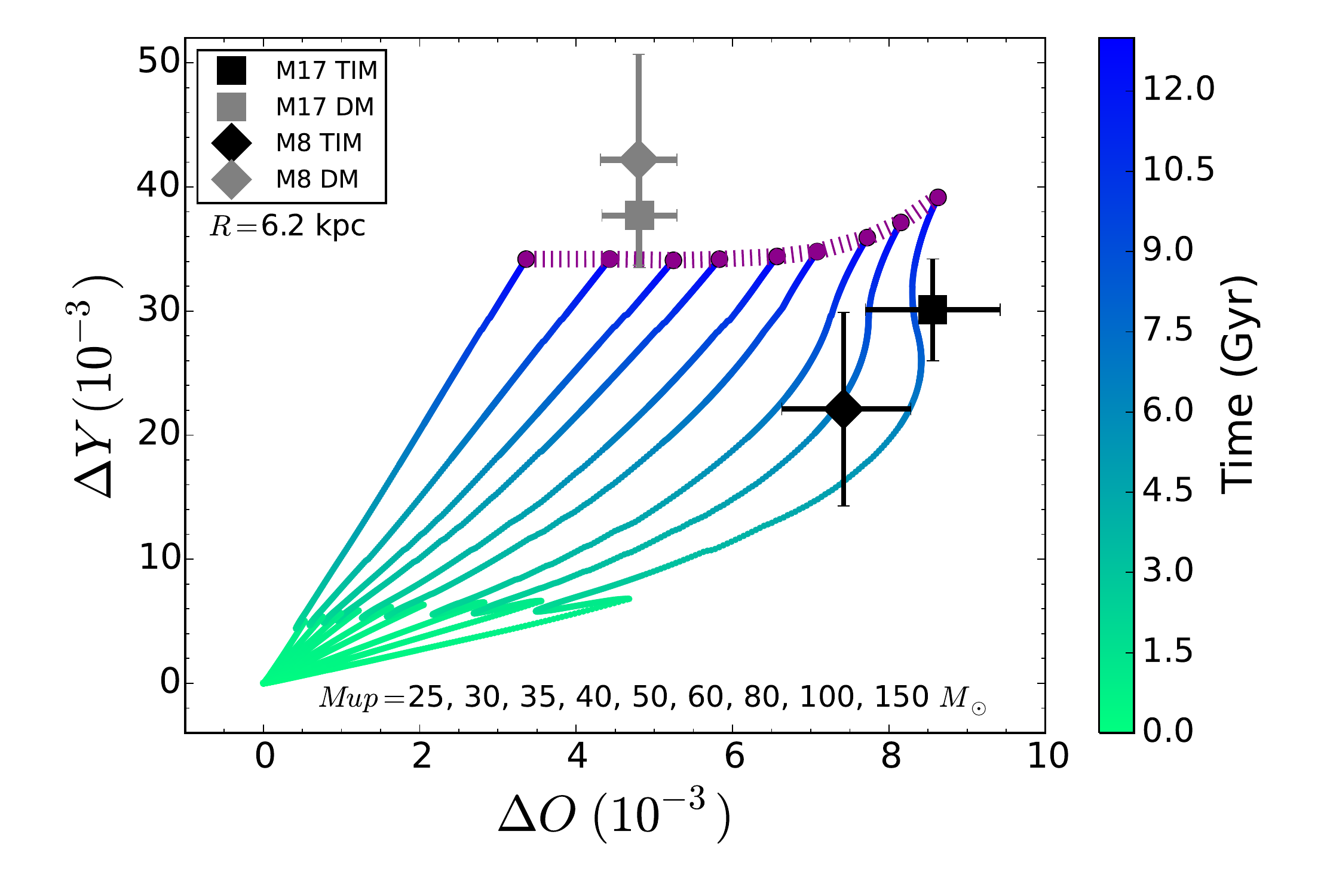}
\end{center}
\caption[f2.eps]{Chemical evolution for $Y$ and $O$ at a galactocentric distance  of 6.2 kpc (approximatelly the distance of M17 and M8). The curves cover the entire evolution from the beginning (0 Gyr) to the present time (13 Gyr,
 magenta points), and each curve corresponds to a model with a different $M_{up}$.  The squares represent the O and He abundances derived for M17 using the DM (grey) and the TIM (black); the diamonds represent the abundances derived for M8 using the DM (grey) and the TIM (black). 
 The dotted magenta line connects the present-time values predicted by the models.
Note that, the observed values should be compared with this magenta line to choose the better $M_{up}$ values.
\label{f-M17-HevsO}}
\end{figure}

\begin{figure}
\begin{center}
\includegraphics[angle=0,scale=0.55]{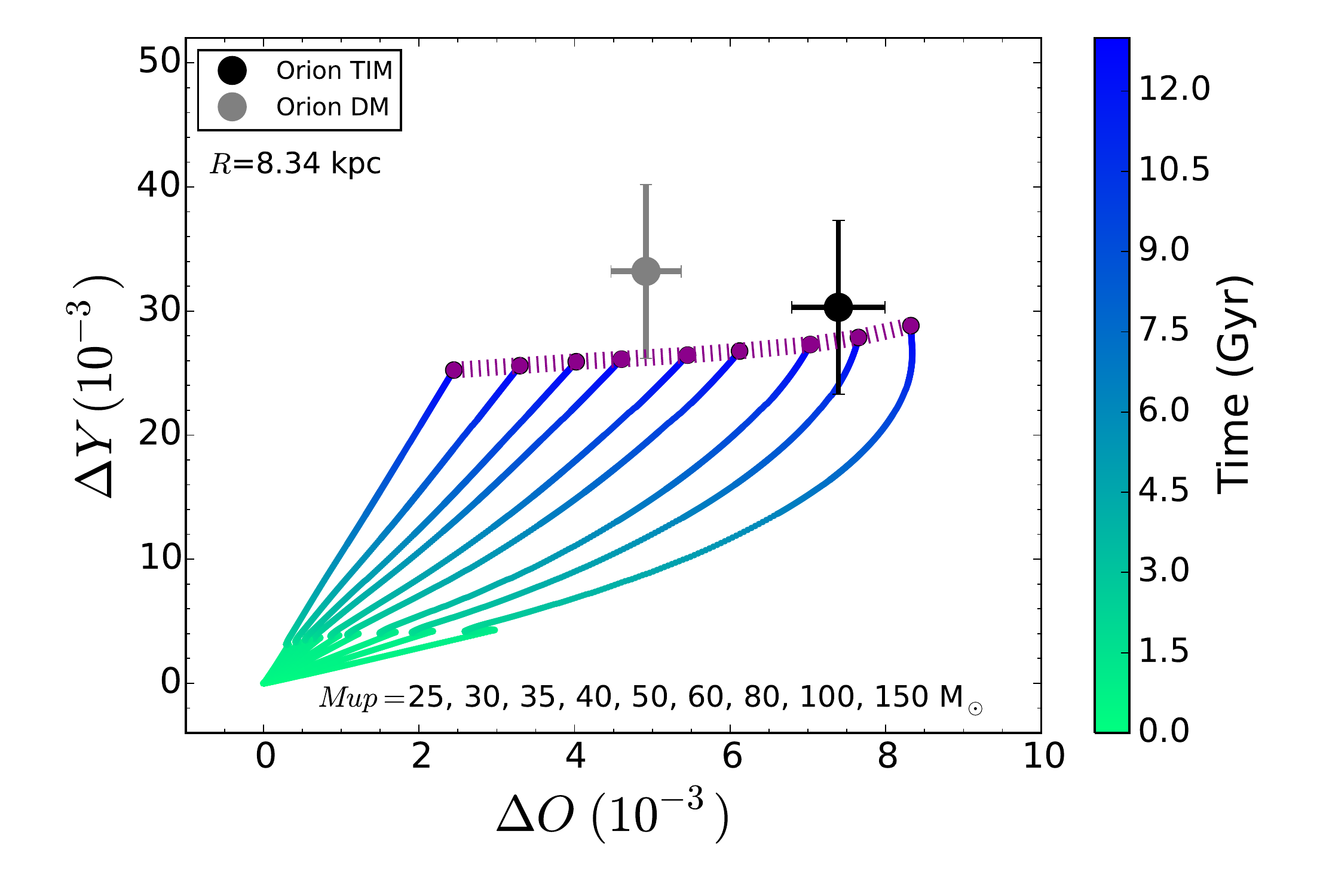}
\end{center}
\caption[f3.eps]{
Chemical evolution for $Y$ and $O$ at a galactocentric distance of 8.34 kpc (Orion). The curves cover the entire evolution from the beginning (0 Gyr) to the present time (13 Gyr,  magenta points), and each curve corresponds to a model with a different $M_{up}$. The circles represent the O and He abundances derived for Orion using the DM (grey) and the TIM (black). 
 Note that, the Orion data should be compared with the magenta line (theoretical present-time values).

\label{f-Orion-HevsO}}
\end{figure}

In Figures~\ref{f-M17-HevsO} and \ref{f-Orion-HevsO} we present the $\Delta O$ vs $\Delta Y$ evolution for $R= 6.2$ and 8.34 kpc, respectively. Moreover, in Figures~\ref{f-M17-OvsC} and \ref{f-Orion-OvsC} we show $\Delta O$ vs $\Delta C$ evolution curves computed for the same radii. The evolution curves are presented for nine CEMs that consider $M_{up}=$ 25, 30, 35, 40, 50, 60, 80, 100, 150 M$_{\odot}$. For comparison, we present observational data, for M17,  M8, and Orion, using the DM and the TIM.

In \citet{carigi19}, models were built to reproduce the radial behavior of the total O/H from 21 \Hiirs. To fit a representative absolute value of the gradient, they inferred two $M_{up}$ values: one if the gaseous O/H values were determined from the DM ($M_{up}=40$ M$_{\odot}$) and the other if the gaseous O/H values were determined from the TIM ($M_{up}=80$ M$_{\odot}$), see Fig.~\ref{f-O/HvsR}.

Also, for NGC 6822 (an irregular galaxy), \citet{hernandez11} built chemical evolution models to reproduce O/H values determined from DM and TIM, and obtained $M_{up}=40$ M$_{\odot}$ and $M_{up}=80$ M$_{\odot}$, respectively, in agreement with the values found for the MW.

In this work, we will obtain uncertainty bars for the $M_{up}$ values, comparing the present-time abundances computed by models using different $M_{up}$ values, with the O/H, He/H, and C/H abundances (and their error bars) for M17,  M8, and Orion.

\subsection{Object selection \label{s-selection}}

The approach in this study is to put quality over quantity;  thus, we only use three objects for our studies: the Orion Nebula, M17,  and M8 since they are the most studied Galactic \Hiirs. One very important characteristic of the Orion Nebula and M17 is that they are very bright (hence their many studies); as a consequence of this they have arguably the best O/H abundance ratio determinations. Another reason to select them, is that they have noticeable different galactocentric radii. While the Orion Nebula is, by far, the best studied \Hiir, M17 has the advantage of being a high ionization \Hiir\ and thus we do not need to worry about an uncertain ICF(He), unfortunately there are no UV observations of M17 (probably due to its relatively high $c({\rm H}\beta)=1.17$) and it is therefore not possible to derive C abundances using CELs and the direct method. We selected M8 because it is also nearby and very bright, its galactocentric distance is very similar to that of M17, allowing us to present both of them using the same simulation and figures, it is the 4th most observed Galactic \Hiir, and probably the 3rd best one suited for a study such as the one we present.

Tables \ref{t-M17obs}, \ref{t-M8obs}, and \ref{t-ORIobs} show the total abundances (gas + dust) by mass derived for M17, M8, and Orion, respectively. The values were derived by transforming the abundances by number obtained by \citet{garcia07, esteban05} for O/H, He/H, and C/H and assuming that $O$ is approximately 45\% of $Z$. The first column of these tables shows the abundances derived through the DM (from CELs) and assumes a constant temperature over the observed volume whereas the second column shows the abundances derived through the TIM (from RLs).

As mentioned above, the line of sight in the direction of M17 has a relatively high reddening, and the $\lambda\lambda$ 1906-1909 \AA\ [\ion{C}{3}] lines are too obscured to have been observed. One might be interested in using the $\lambda$ 4267 \AA\ \ion{C}{2} to complete the DM determination; $\lambda$ 4267 \AA\ \ion{C}{2}, can not be used as part as the DM for the same reasons that the $\lambda$ 4650 \AA\ \ion{O}{2} multiplet can not be used (since it corresponds to the TIM). Therefore, while widely used, it should not be considered as part of the DM (and one should beware of authors that use it as part of the DM without a clear and consistent explanation on the ADF origin and its consequences).

\begin{table}[!t]\centering
\setlength{\tabnotewidth}{7.2cm}
\tablecols{3}
\setlength{\tabcolsep}{2\tabcolsep}
\caption{M17: Observed $\Delta Y$, \zo , and \zc\ values\TA} \label{t-M17obs}
\begin{tabular}{ccc}
\toprule
& \multicolumn{1}{c}{DM} & \multicolumn{1}{c}{TIM} \\
\midrule
$\Delta Y\,(10^{-3})$  & 37.7$\pm$4.2   & 30.1$\pm$4.1  \\
\zo $\,(10^{-3})$      & 4.81$\pm$0.48  & 8.56$\pm$0.86 \\
\zc $\,(10^{-3})$      & ...            & 6.27$\pm$0.63 \\ 
$\Delta Y/\Delta O$    & 7.86$\pm$1.39  & 3.52$\pm$0.72 \\ 
$\Delta C/\Delta O$    & ...            & 0.73$\pm$0.10 \\  
\bottomrule
\tabnotetext{a}{Observations from \citet{garcia07}.}
\end{tabular}
\end{table}

\begin{table}[!t]\centering
\setlength{\tabnotewidth}{7.2cm}
\tablecols{3}
\setlength{\tabcolsep}{2\tabcolsep}
\caption{M8: Observed $\Delta Y$, \zo , and \zc\ values\TA} \label{t-M8obs}
\begin{tabular}{ccc}
\toprule
& \multicolumn{1}{c}{DM} & \multicolumn{1}{c}{TIM} \\
\midrule
$\Delta Y\,(10^{-3})$  & 42.2$\pm$8.5           & 22.1$\pm$7.8           \\
\zo $\,(10^{-3})$      & 4.80$\pm$0.49          & 7.42$\pm$0.79          \\
\zc $\,(10^{-3})$      & 1.50$^{+0.30}_{-0.75}$ & 5.30$^{+1.07}_{-2.65}$ \\ 
$\Delta Y/\Delta O$    & 8.80$\pm$1.99          & 2.84$\pm$1.06          \\ 
$\Delta C/\Delta O$    & 0.31$^{+0.07}_{-0.16}$ & 0.68$^{+0.15}_{-0.35}$ \\  
\bottomrule
\tabnotetext{a}{Observations from \citet{garcia07}.}
\end{tabular}
\end{table}

\begin{table}[!t]\centering
\setlength{\tabnotewidth}{7.2cm}
\tablecols{3}
\setlength{\tabcolsep}{2\tabcolsep}
\caption{Orion Nebula: Observed $\Delta Y$, \zo , and \zc\ values\TA} \label{t-ORIobs}
 \begin{tabular}{lcc}
    \toprule
     & \multicolumn{1}{c}{DM} & \multicolumn{1}{c}{TIM} \\
    \midrule
$\Delta Y\,(10^{-3})$  & 33.2$\pm$7.1  &  30.3$\pm$7.0  \\
\zo $\,(10^{-3})$      & 4.92$\pm$0.45 &  7.39$\pm$0.60 \\
\zc $\,(10^{-3})$      & 1.35$^{+0.55}_{-0.40}$ & 2.81$^{+0.42}_{-0.32}$ \\ 
$\Delta Y/\Delta O$    & 6.75$\pm$1.45 & 4.10$\pm$1.01 \\ 
$\Delta C/\Delta O$    & 0.27$^{+0.12}_{-0.08}$ & 0.38$^{+0.06}_{-0.05}$ \\
    \bottomrule
    \tabnotetext{a}{Observations from \citet{esteban05}.}
  \end{tabular}
\end{table}

\subsection{O/H vs. He/H}

Gaseous O/H abundances are readily available from many observational sources, however they are frequently not converted to the total ISM O/H; to do this, it is necessary to include the fraction of O trapped in dust grains. It is estimated that this correction is  between 0.07 and 0.13 dex for most \Hiirs\ \citep{mesa09,peimbert10,espiritu17}; the exact value depends on the metallicity and on the efficiency of the dust destruction present within each \Hiir.
Here we will include a correction of 0.12 dex for the Orion Nebula \citep{mesa09,espiritu17} and 0.11 dex for both M17 and M8\citep{peimbert10}. 

The C/H abundance should also be corrected for dust depletion; this correction is expected to be similar or slightly lower than the O/H correction \citep{esteban98,esteban09}. Here we will assume a correction of 0.10 dex for all three \Hiirs.

Although most elements should include a correction due to dust depletion, the fact that He is an inert noble gas means that no correction will be necessary for the He/H abundances. 

\subsubsection{\zo\ vs. $Y$ for M17  and M8}

In Figure~\ref{f-M17-HevsO} we present the theoretical evolution of $\Delta O$ and $\Delta Y$ vs time, for $R=6.2$ kpc. We plot nine curves that correspond to the nine $M_{up}$ values listed in section 2.2. The curves begin at $t=0$ Gyr ($\Delta O = \Delta Y=0$) and end at 13 Gyr. We include the $\Delta O$  and $\Delta Y$ values for M17 and M8, determined from the DM and the TIM (see Tables~\ref{t-M17obs} and \ref{t-M8obs}). In order to choose the $M_{up}$ values that better reproduce the observational data, the top of each curve (the predicted values at present time (shown in magenta points) should be compared with the M17 and M8 data (the observed abundances for the ISM).

\begin{table}[!t]\centering
  \setlength{\tabnotewidth}{\columnwidth}
  \tablecols{6}
  \caption{Present day values in the ISM predicted by the models for  $R=6.2$ kpc (M17 and M8).} \label{t-M17mod}
 \begin{tabular}{lccccc}
    \toprule
   \multicolumn{1}{c}{$M_{up}$} & \multicolumn{1}{c}{$\Delta Y$ ($10^{-3})$} & \multicolumn{1}{c}{\zo ($10^{-3}$)} & \multicolumn{1}{c}{\zc ($10^{-3}$)} & \multicolumn{1}{c}{$\Delta Y/\Delta O$} & \multicolumn{1}{c}{$\Delta Y/\Delta C$} \\
    \midrule
    150 & 39.16 & 8.63 & 7.54 & 4.54 & 1.14 \\
    100 & 37.14 & 8.15 & 7.08 & 4.56 & 1.15 \\ 
   80   & 35.92 & 7.72 & 6.67 & 4.65 & 1.16 \\
   60   & 34.79 & 7.08 & 6.01 & 4.91 & 1.18 \\
   50   & 34.40 & 6.57 & 5.47 & 5.24 & 1.20 \\
   40   & 34.18 & 5.83 & 4.86 & 5.86 & 1.20 \\
   35   & 34.09 & 5.24 & 4.50 & 6.51 & 1.16 \\
   30   & 34.19 & 4.43 & 4.28 & 7.72 & 1.04 \\
   25   & 34.19 & 3.36 & 4.24 & 8.06 & 0.79 \\
    \bottomrule
  \end{tabular}
\end{table}

The curves evolve more rapidly to the right with increasing $M_{up}$ values, because the O production for high mass stars (HMS) increases with the stellar mass. At 1 Gyr, when the halo formation ends, the curves present a loop due to the dilution of the ISM with primordial infall \citep[$Y=0.2451$, $O=0.0$][]{valerdi19} that forms the disk. The rest of the evolution depends on the lifetime and the initial metallicity ($Z$) of the HMS and low-and-intermediate mass stars (LIMS), see \citet{carigi08} and \citet{carigi11} .

Current $Y$ values are almost constant for $M_{up} < 50$ M$_{\odot}$ and increase for $M_{up} > 50$ M$_{\odot}$  (see Table~\ref{t-M17mod}), because: i) for low $Z$ (equivalent to low $O$), HMS in the 8-25 ${\rm M}_{\odot}$ range are much more efficient to produce He than those in the 25-150 ${\rm M}_{\odot}$ range, ii) for high $Z$, HMS in the 50-150 ${\rm M}_{\odot}$ range are very efficient to produce He, and iii) for LIMS, the He contribution is not strongly $Z$-dependent. 

A peculiarity of Figure~\ref{f-M17-HevsO} is the shape of the $M_{up}=150\;{\rm M}_{\odot}$  curve, where the oxygen abundance diminishes between 7.8 and 10.0 Gyrs. This O dilution is caused by the huge amount of C ejected by very massive stars of high $Z$ (see the description of the carbon evolution in Section 2.5.1).

When comparing the observed O and He values for M17 with those derived from our models we find: i)~for the $\Delta O$ value determined with the DM, an IMF with a galactic $M_{up}$ of $30\;$-$\; 36 \ {\rm M}_{\odot}$, while ii)~for the $\Delta O$ value determined with the TIM, an $M_{up} > 75 \ {\rm M}_{\odot}$, iii)~for the $\Delta Y$ value determined with the DM, all values of $M_{up}$ are allowed (the $\Delta Y$ value is not very restrictive), and iv)
~for the $\Delta Y$ value determined with the TIM, an $M_{up} < 70 \ {\rm M}_{\odot} (1 \sigma)$ (allowing for all possible values at the $2 \sigma$ level).

From the M8 values we find: i)~the $\Delta O$ value determined with the DM is nearly identical to the one determined for M17 therefore the range determined for $M_{up}$ is also of $30\;$-$\; 36 \ {\rm M}_{\odot}$, ii)~the $\Delta O$ value determined with the TIM suggests $M_{up}$ in the $52\;$-$\; 120 \ {\rm M}_{\odot}$ range, iii)~for the $\Delta Y$ value determined with the DM, all values of $M_{up}$ are allowed (the $\Delta Y$ value is not very restrictive), and iv)~for the $\Delta Y$ value determined with the TIM there is no solution at the $1\sigma$ level, yet at the  $2 \sigma$ level all solutions are allowed, this shows both: the uncertainty of the $ICF({\rm He})$ and the lack of restriction produced by $\Delta Y$.

\subsubsection{\zo\ vs. $Y$ for the Orion Nebula}

In Figure~\ref{f-Orion-HevsO} we present curves of theoretical evolution of $\Delta O$ and $\Delta Y$ vs time for $R=8.34$ kpc, corresponding to nine $M_{up}$ values listed in section 2.2. Moreover, we include the $\Delta O$ and $\Delta Y$ values for Orion, determined from the DM and the TIM (see Table~\ref{t-ORIobs}). As in Fig.~\ref{f-M17-HevsO}, the top of each curve (in magenta)
corresponds to the end of evolution (i.e. the present time; see columns 2 and 3 of Table~\ref{t-ORImod}), and should be compared with the Orion data, to choose the $M_{up}$ values that better reproduce the observations.

\begin{table}[!t]\centering
  \setlength{\tabnotewidth}{\columnwidth}
  \tablecols{6}
  \caption{Present day values in the ISM predicted by the models for $R=8.34$ kpc (Orion Nebula)} \label{t-ORImod}
 \begin{tabular}{lccccc}
    \toprule
   \multicolumn{1}{c}{$M_{up}$} & \multicolumn{1}{c}{$\Delta Y$($10^{-3}$)} & \multicolumn{1}{c}{\zo ($10^{-3}$)} & \multicolumn{1}{c}{\zc ($10^{-3}$)} & 
    \multicolumn{1}{c}{$\Delta Y/\Delta O$}&\multicolumn{1}{c}{$\Delta Y/\Delta C$} \\
    \midrule
150 & 28.82 & 8.32 & 5.22 & 3.46 & 5.52 \\
100 & 27.87 & 7.65 & 4.84 & 3.64 & 5.76 \\
80   & 27.30 & 7.03 & 4.54 & 3.88 & 6.01 \\
60   & 26.77 & 6.12 & 4.10 & 4.37 & 6.53 \\
50   & 26.45 & 5.45 & 3.77 & 4.85 & 7.02 \\
40   & 26.12 & 4.60 & 3.42 & 5.68 & 7.64 \\
35   & 25.91 & 4.02 & 3.24 & 6.45 & 8.00 \\
30   & 25.50 & 3.30 & 3.11 & 7.73 & 8.20 \\
25   & 25.24 & 2.45 & 3.04 & 10.30 & 8.30 \\
    \bottomrule
  \end{tabular}
\end{table}

In Figure~\ref{f-Orion-HevsO}, for any given $M_{up}$, the evolutionary curves reach lower $\Delta O$ and $\Delta Y$ values than the corresponding coeval values in Fig. 2, because the O/H gradient is negative for all MW-like models. Therefore, for any given time, the $\Delta O$ value (and $Z$ value) for $R=8.34$ kpc is lower than $\Delta O$ value for $R=6.2$ kpc. Consequently, very few massive stars of high $Z$ form and, since they are more efficient He producers, the reached $Y$ values are lower. In this figure, for the $M_{up}=150$ curve, the O dilution is lower, due to the relative lack of massive stars of high $Z$ (high-efficient C producers).

When comparing the observed O and He values for Orion with those derived from our models, we find: i)~for the $\Delta O$ value determined with the DM, an $M_{up}$ of $38\;$-$\; 50 \ {\rm M}_{\odot}$, while ii)~for the $\Delta O$ value determined with the TIM, an $M_{up}$ of $75\;$-$\; 130 \ {\rm M}_{\odot}$, iii)~for the $\Delta Y$ value determined with the DM, an $M_{up} > 35 \ {\rm M}_{\odot} (1 \sigma)$ (allowing for all possible values at the $1.2 \sigma$ level), and iv)~for the $\Delta Y$ value from the TIM, again, all values of $M_{up}$ are allowed.

\subsection{O vs. C}

\subsubsection{O vs. C for M17  and M8}

\begin{figure}
\begin{center}
\includegraphics[angle=0,scale=0.55]{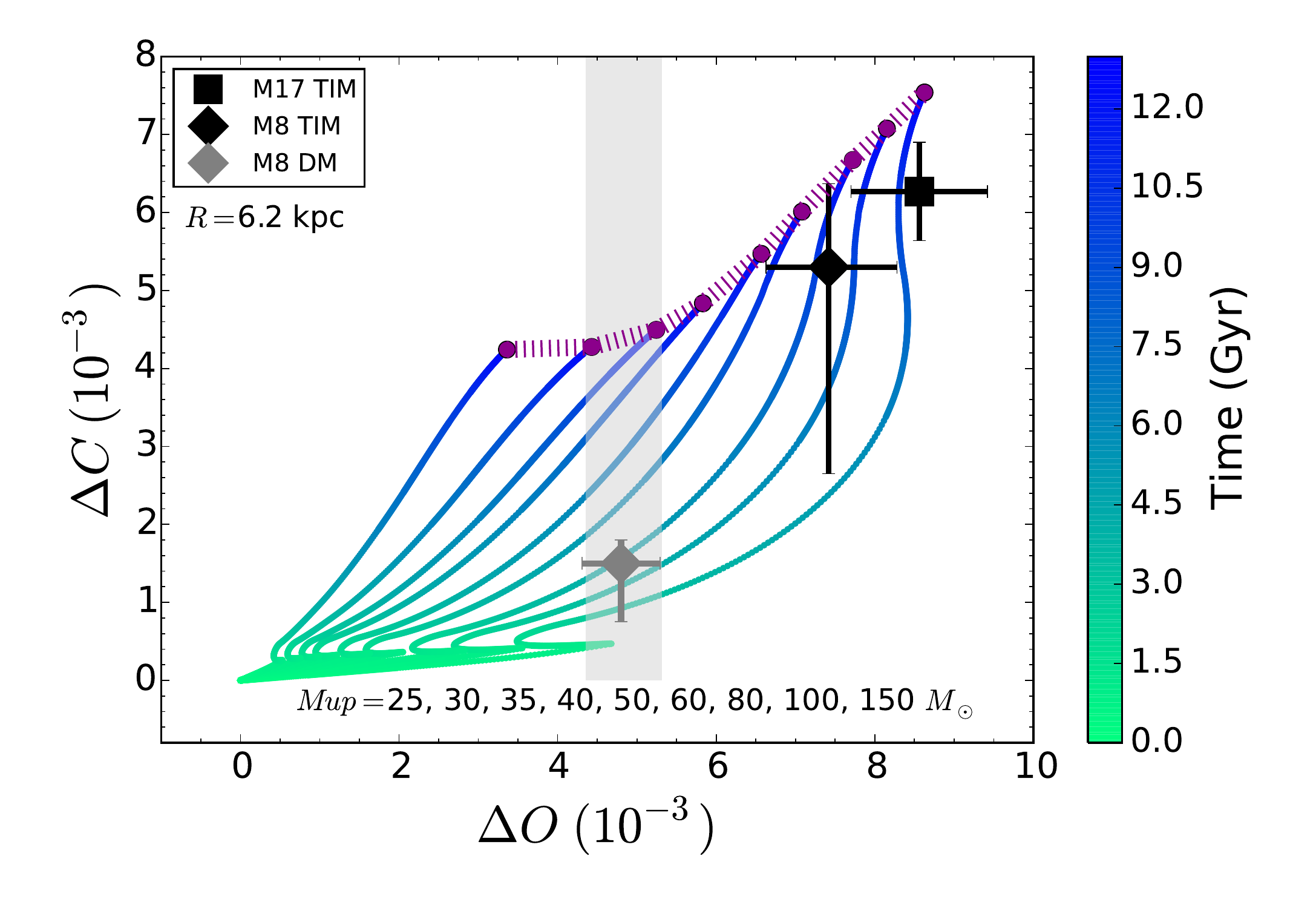}
\end{center}
\caption[f4.eps]{
Chemical evolution for $C$ and $O$ at a  galactocentric distance of 6.2 kpc (approximatelly the distance of M17 and M8). The curves and  points are similar to those in Fig. 2.  Notice however, there are only 3 points in this figure instead of the 4 in Fig. 2: unfortunately there is no restriction on $C$ for the DM of M17 and, instead of the fourth point, the shaded vertical band is the $\Delta O$ predicted by the DM. Again, the observed values should be compared with the magenta curve.
\label{f-M17-OvsC}}
\end{figure}

In Figure~\ref{f-M17-OvsC} we show the curves of theoretical evolution of $\Delta O$ and $\Delta C$ vs time, for $R=6.2$ kpc, obtained from our nine CEMs. The current values, at the top end of the curves, are presented in columns 3 and 4 of Table~\ref{t-M17mod}. We include the observed $\Delta O$ and $\Delta C$ values for M17 and M8 determined from the TIM.
The gray diamond represents the $\Delta O$ and $\Delta C$ values determined for M8, while the shaded vertical bar represents the $\Delta O$ for M17 using the DM combined with the lack of a C determination available from the DM (see Table~\ref{t-M17obs}). 

One note on the $\Delta C$ determination for M8: according to the recent ICFs computed for giant \Hiirs\ (Amayo, Delgado-Inglada, Stasi\'nska, 2020, in prep.), the use of C/O = C$^{++}$/O$^{++}$ in M8 may underestimate real C/O value by up to $\sim0.3$ dex. However, these ICFs  may not be adequate for Galactic \Hiirs\ where only a small area is observed. We decided not to change the value of C/H but to increase the associated error bars.

Current $C$ values are almost constant for $M_{up} \lesssim 35$ M$_{\odot}$ and increase significantly for $M_{up} \gtrsim 40$ M$_{\odot}$ (see Table~\ref{t-M17mod}); this happens for two main reasons: i) stars in the 40-150 ${\rm M}_{\odot}$ range produce much more C when they are more metal rich, and ii) stars in the 1-3 ${\rm M}_{\odot}$ range produce more C when they are metal poor. Due to the the LMS enrichment contribution, HMS of high $Z$ contribute at similar times than LMS of low $Z$ \citep[e.g.][]{akerman04,carigi05,carigi11}.

When comparing the observed $\Delta C$ value for M17 determined with the TIM  with those derived from the CEMs, we find an $M_{up}$ in the $55 \;$-$\; 95 \ {\rm M}_{\odot}$ range.
Using the determination from the M8's $\Delta C$ measurements obtained with the TIM, we only find an upper limit $M_{up}< 72 \ {\rm M}_{\odot}$, the lack of a lower limit is due to the large uncertainty on the lower limit of the ICF. When determining $\Delta C$ value for M8 using the DM the value is not compatible with our models, with the DM falling short of our models by a factor of at least 2.5 (for the lowest $M_{up}= 30 \ {\rm M}_{\odot}$), and probably a factor of about 3.5 or more (for a more reasonable $M_{up} \gtrsim 70 \ {\rm M}_{\odot}$).  Finaly, regarding the DM determination for M17: due to the high reddening of M17, it has not been observed in the UV, and therefore it has not been possible to obtain [\ion{C}{3}] intensities, nor to determine DM abundances. The $\Delta O$ $M_{up}$ determinations are the same as those in Fig.~\ref{f-M17-HevsO}.

\subsubsection{O vs. C for Orion}

\begin{figure}
\begin{center}
\includegraphics[angle=0,scale=0.55]{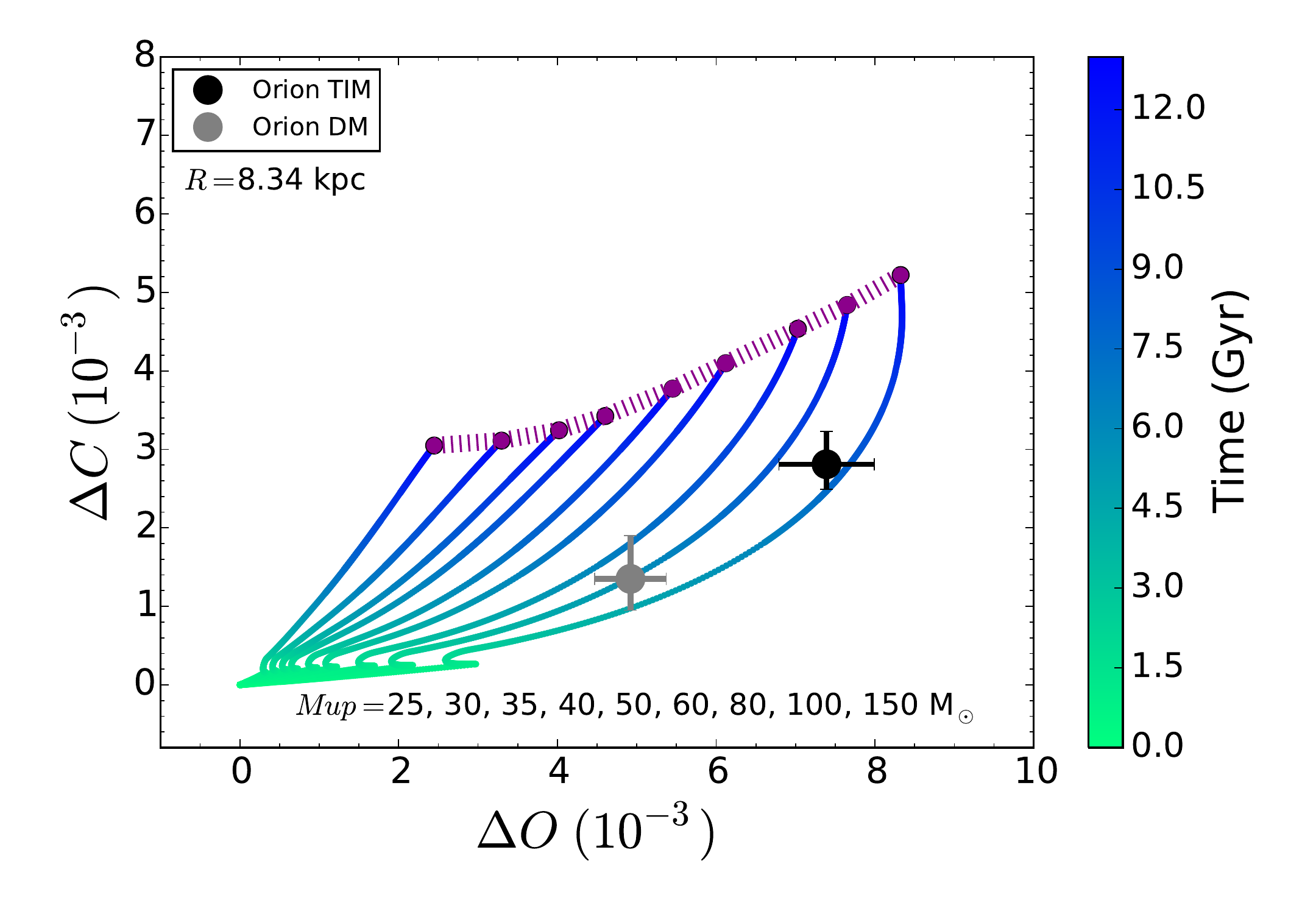}
\end{center}
\caption[f5.eps]{
Chemical evolution for $C$ and $O$ for a galactocentric distance of 8.34 kpc (Orion). The curves and points are similar to those in Fig. 3.  The Orion data should be compared with the magenta line.
\label{f-Orion-OvsC}}
\end{figure}

In Figure~\ref{f-Orion-OvsC} we show the theoretical evolution of $\Delta O$ and $\Delta C$ vs time for $R=8.34$ kpc, for each of our nine $M_{up}$ values. Moreover, we include the $\Delta O$ and $\Delta C$ values for Orion, determined from the DM and the TIM (see Table~\ref{t-ORIobs}). In this figure, for any given $M_{up}$, the evolutionary curves reach lower $O$ and $C$ values than the coeval values in  Fig.~\ref{f-M17-OvsC}, because the O/H and C/H gradients are negative for all the MW-like models.

By comparing the observed C values for Orion with the current values derived from our models (see Table~\ref{t-ORImod}), we find that neither set of observed $\Delta C$ abundances (neither DM nor TIM) are consistent with the theoretical predictions. Since the $\Delta O$ values and determinations are the same as those in Fig.~\ref{f-Orion-HevsO} it seems that the C abundance in Orion is lower than expected; however, the C/H abundance in the Orion nebula is noticeable smaller (about 0.1 dex) than in NGC~3603 ($R=8.65\,$kpc) and NGC~3576 ($R=7.46\,$kpc), which are the two \Hiirs\ with a galactocentric distance closest to the one of the Orion nebula \citep{esteban04,garcia04,garcia06}. There are two causes that may explain this low value. The first one is the use of an inadequate ionization correction factor (ICF) but according to a recent study on ICFs for giant \Hii\ regions (Amayo et al. (2020) in prep.) it is adequate to use C/O = C$^{++}$/O$^{++}$ to compute C abundances. The second one is the possibility of this nebula having more C atoms deposited in dust grains and thus, a lower gaseous abundance of C; a clue that the dust properties in Orion are different than in other \Hiirs\ is the high total to selective absorption ratio present in Orion, that amounts to about $R_{V}=E_V/E_{B-V}\approx5.5$ \citep{peimbert69,esteban04}, while for most other objects it amounts to about $R_V=3.1$ \citep{cardelli89}. If we consider that Orion could have a slightly higher (0.1 dex) total C abundance, we find, for the TIM, the total gas plus dust ratio of C/H  for the Orion nebula to be $12 + \log ({\rm C/H}) \approx 8.53 \pm 0.08$: this value represents a slightly lower $M_{up}$ than the one derived from O/H. On the other hand, we find, for the DM, the total C/H to be $12 + \log ({\rm C/H}) \approx 8.22 \pm 0.10$, still 0.1 dex less than our lowest model, approximately 0.2 dex lower than the $M_{up}\approx 40$ \msun\ favored by the O/H determinations of the DM, and approximately 0.3 dex lower than what is required to be consistent with the more favored $M_{up} \lesssim 80$ \msun\ values.

\section{The chemistry has better memory than the light}

The lifetime of massive stars (few Myr) is very short compared to the age of galaxies (several Gyr); consequently massive stars are not an important component of the light of the majority of the observed galaxies; yet, when massive stars die, their contributions are quickly incorporated in the chemistry of the ISM. 

The chemical composition of an \Hiir\ is the result of the whole history of the chemical evolution of any given galaxy; therefore the chemical composition of \Hiirs\ can be compared with estimates of the present day chemistry derived from galactic CEMs; and we can infer the amount of formed massive stars (equivalently $M_{up}$ value), comparing the chemical abundances in the ISM with those obtained from CEMs.

To determine the most massive stars in the MW by looking for them at present has several major inconveniences: there are very few of these stars, this is compounded by the fact that they must be formed in very massive giant molecular clouds (GMCs), and they are the first ones to evolve (first shedding mass in strong stellar winds, and then going supernova, all this before the GMC is dissipated by the combined effect of the stars that are evolving inside it), thus they are usually obscured and very difficult to observe during their short lifespan. Moreover, most massive stars that can be observed today, may not be representative of the most massive stars that have existed during the evolution of the Galaxy, i.e. the stars that have contributed to the evolution of the chemistry of the present day ISM as well as the chemistry available during the most recent star formation.

Based on the $\Delta Y$ comparison, between the observed values (from TIM or DM) and the preset-time predicted values (from CEMs), we cannot exclude any $M_{up}$ value in the 25 - 150 \msun\ range. However, from the $\Delta O$ comparison, we find agreement for $M_{up}$ values  in the $30 < M_{up} < 50$ \msun\ range for the DM, and in the $75 < M_{up} < 120$ \msun\ range for the TIM. Moreover, from the $\Delta C$ comparison, we find agreement in the $55 < M_{up} < 95$ \msun\ range for the TIM, and a suggestion of a very small $M_{up}$ for the DM ($M_{up} < 25$ \msun). Therefore, by comparing TIM with CEMs, the best $M_{up}$ values are in the $70 \lesssim M_{up} \lesssim 100$ \msun\ range; and by comparing DM with CEMs, the best $M_{up}$ values are in the $25 \lesssim M_{up} \lesssim 45$ \msun range.

\citet{weidner13} in their figure 3 showed the dependence of the star formation rate (SFR) on the integrated galactic stellar initial mass function (IGIMF, called IMF in our CEMs) for different power-law indexes, ($\alpha$, for initial stellar masses between 1.3 \msun\ and  $M_{up}$). They noted that IMF for a ${\rm SFR} \approx 1$ \msun/yr corresponds to an $\alpha=2.6$ and to an $M_{up} \approx$ 100 \msun; and that these values are in agreement with the MW (we use $\alpha=2.7$ for our CEMs). On the other hand the SFR required to obtain $M_{up} \approx 40$ \msun\ is ${\rm SFR} \approx 10^{-2}$ \msun/yr (as well as an $\alpha \approx 2.8$).

Regarding the observational determinations of the SFR: i) the MW, a spiral galaxy (Sbc) with total stellar mass $ \sim 10^{11}$ \msun, presents a galaxy-wide SFR $\sim 0.7-2.3$ \msun/yr \citep{robi10,chomiuk11}; ii) NGC 300, a small spiral galaxy (Sd) with total stellar mass $= 1.9 \times 10^9$ \msun,  presents a galaxy-wide SFR $= 0.08 - 0.30$ \msun/yr, approximately an order of magnitude smaller than the SFR in the MW \citep[e.g.][and references therein]{kang16}.

Moreover, based on spatially-resolved spectroscopic properties of low-redshift star-forming galaxies, \citet{sanchez19} showed in his Fig.~7 a difference of approximately 1 order of magnitude between the SFR of Sbc galaxies with stellar mass $\sim 10^{11}$ \msun \ (as the MW galaxy),  and the SFR of Sd galaxies with stellar mass $\sim 2 \times 10^{9}$ \msun \ (as NGC~300).

Therefore, the $M_{up}$ values we derive from the TIM are consistent with the SFR of the MW galaxy, while the $M_{up}$ values derived from the DM are consistent with a galaxy with a mass and SFR similar to those of NGC~300, but not with the mass and SFR of the MW.

Moreover the abundances derived from the TIM are consistent with those derived from observation of other young objects in the MW (Cepheids, B stars), while abundances derived from the DM are approximately 0.25 dex too small.

The chemical composition of a given \Hiir\  is the result of the evolution of the ISM throughout the history of our Galaxy, therefore the chemical abundances of an \Hiir\ do not depend on the IMF of the observed \Hiir.  In particular the most massive star of a given \Hiir\ is not representative of the most massive stars of the galactic IMF.

The chemical composition of a given \Hiir\ is the result of the chemical evolution of the galaxy in question.  Therefore the chemical composition of a given \Hiir\ is the result of the evolution of the galaxy during its lifetime. The  IMF during the evolution of the galaxy provides us with the expected chemical abundances for a given \Hiir.

\section{Conclusions}

We have computed nine chemical evolution models (CEM) of a MW like galaxy, the only difference among these models is the IMF, specifically its $M_{\rm up}$ value, that ranges between 25 and 150 \msun. We compare the model predictions with the O/H, He/H, and C/H values derived for three Galactic \Hiirs : M17, M8, and Orion. We computed the abundances by two different methods: the DM (direct method) and the TIM (temperature independent method); these methods have always given  different results. We selected these objects because they  probably have the best O/H determinations and because their galactocentric radii are different enough to be useful as independent constraints for the CEMs (6.2 and 8.34 kpc, respectively). The comparison between models and observations tells us which is the $M_{up}$ that better fits each set of observations.

It is useful to remember that the chemistry has better memory than the observed UV light. In other words: the chemistry will explore the average $M_{up}$ over the lifespan of the MW, while any measurement of the UV radiation or of the most massive stars observed can only be a reflection of the present day $M_{up}$ (and can potentially have significant biases toward lower masses).

When comparing the models with the DM abundances, we find: for $\Delta O$, $30 < M_{up} < 50$ \msun\ range; for $\Delta C$, $M_{up} < 25$ \msun\ ; while $Y$, does not represent a significant restriction. Overall the DM produces a preferred values in the $25 < M_{up} < 45$ \msun\ range. On the other hand, when using the TIM abundances we find: for $\Delta O$, $52 < M_{up} < 150$ \msun\ range; for $\Delta C$, $25 < M_{up} < 95$ \msun\ range; while $Y$, suggests a smaller value, but does not represent a significant restriction. Overall the TIM produces a preferred values in the $70 < M_{up} < 100$ \msun\ range.

Moreover the $M_{up}$ in a given galaxy is directly related to the SFR, and the SFR is directly related to the mass of any given galaxy. A MW like galaxy, with a SFR~$\sim$~1~\msun/yr, is expected to have an $M_{up} \approx 100$ \msun; in good agreement with the TIM determination, but not with the DM determination (which would be more consistent with an Sd galaxy with a SFR~$\sim$~0.01~\msun/yr).

\

We are grateful to the referee for excellent suggestions.
The authors acknowledge support from PAPIIT (DGAPA-UNAM), grants No. IA-100420, IA-101517, IG-100319, IN-100519, and IN-103820.
L.C. is thankful for the financial support by CONACYT grant FC-2016-01-1916.

\clearpage

\end{document}